\documentclass[showpacs,aps,graphicx,twocolumn]{revtex4}
\usepackage{graphicx}

\begin{document}

\title{ One-step  deterministic polarization entanglement purification using spatial entanglement\footnote{Published in Phys. Rev. A \textbf{82}, 044305 (2010)}}

\author{Yu-Bo Sheng$^{1,2}$  and Fu-Guo Deng$^{1}$\footnote{Corresponding author: fgdeng@bnu.edu.cn}}
\address{$^1$ Department of Physics, Beijing Normal University, Beijing 100875, China\\
$^2$ Key Laboratory of Beam Technology and Material Modification of
Ministry of Education, College of Nuclear Science and Technology,
Beijing Normal University, Beijing 100875, China }
\date{\today }

\begin{abstract}

We present a one-step deterministic entanglement purification
protocol with linear optics and postselection. Compared with the
Simon-Pan protocol (C. Simon and J. W. Pan, Phys. Rev. Lett.
\textbf{89}, 257901 (2002)), this one-step protocol has some
advantages. First, it can obtain a maximally entangled pair with
only one step, instead of improving the fidelity of less-entangled
photon pairs by performing the protocol repeatedly in other
protocols. Second, it works in a deterministic way, not a
probabilistic one, which greatly reduces the number of entanglement
resources needed. Third, it does not require the polarization state
be entangled;  only spatial entanglement is needed. Moreover, it is
feasible with current techniques (Nature \textbf{423}, 417 (2003)).
All these advantages   make this one-step protocol more convenient
than others in quantum-communication applications.
\end{abstract}
 \maketitle

Entanglement is of vital importance for quantum communication and
computation, especially the distribution of entanglement between
distant locations
\cite{computation1,computation2,teleportation,cteleportation,Ekert91,BBM92,rmp}.
However, entangled quantum systems   inevitably suffer from  channel
noise when the entangled photons propagate away from each other. The
channel noise can be caused by  thermal fluctuation, vibration, the
imperfection of the fiber, and so on. Thus, a maximally entangled
photon pair may become a less entangled pair. That is, it is in a
mixed state. Entanglement purification protocols
\cite{Bennett1,Deutsch,Pan1,Simon,shengpra} are essentially used to
get some maximally entangled photon pairs from a set of
less-entangled photon pairs with the help of local operations and
classical communications. Now, it has become one of the most
important stages in quantum repeaters protocol \cite{repeater} and
long-distance communication protocols \cite{DLCZ,zhao}.

The first entanglement purification protocol (EPP) was proposed by
Bennett \emph{et al}. in 1996 \cite{Bennett1}. Their protocol is
based on controlled-NOT (CNOT) gates and is used to purify a special
state, called a Werner state \cite{werner}. In the same year, Deutsh
\emph{et al}. optimized the first EPP with two additional specific
unitary operations \cite{Deutsch}. With the same logical operation,
Murao \emph{et al}. has proposed an EPP  for purifying multipartite
quantum systems in a Greenberger-Horne-Zeilinger  state \cite{
Murao}. Purification of high-dimension qubit protocols also have
been proposed with  similar quantum logical operations \cite{
Horodecki,Yong}. However, it is very difficult to implement a
perfect CNOT gate with linear optics in current experiments. In
2001, Pan \emph{et al}. proposed an EPP with linear optical elements
and an ideal entanglement source \cite{Pan1}, calling it the
polarizing beam splitter (PBS) protocol. The PBS protocol was
demonstrated in 2003 \cite{Pan2}. In 2002, Simon and Pan proposed
another polarization EPP using spatial entanglement \cite{Simon},
calling it the Simon-Pan protocol. In this EPP, they use the
currently available parametric down-conversion (PDC) source to
substitute the ideal single-pair entanglement source. In 2008, an
EPP using cross-Kerr nonlinearity \cite{shengpra} was proposed and
it is used to purify the entangled photon pairs from a PDC source.
In this protocol, the cross-Kerr nonlinearity is used to construct a
quantum nondemolition detector (QND) for parity-check measurements.
It can be repeated to get a high-fidelity entangled photon pairs
from a practical entanglement source.

Although there are some important EPPs
\cite{Bennett1,Deutsch,Pan1,Simon,shengpra}, none of the
conventional entanglement purification protocols (CEPPs) can,
however, actually obtain a maximally entangled photon pairs
perfectly. They can only increase the fidelity of the entangled
photon pairs in an ensemble in a mixed state by consuming largely
less-entangled photon pairs. For instance, after each purification
step, the new fidelity $F'$ and the initial fidelity $F$ satisfy the
following relation \cite{Bennett1},
\begin{eqnarray}
F'=\frac{F^{2}+\frac{1}{9}(1-F)^{2}}{F^{2}+\frac{2}{3}F(1-F)+\frac{5}{9}(1-F)^{2}}.
\end{eqnarray}
One can easily find that $F'$ increases when $F>\frac{1}{2}$, but
$F=1$ is the local attractor and it cannot actually be reached.
Recently, a deterministic polarization entanglement purification
protocol (DEPP) has been proposed with hyperentanglement
\cite{twostepepp}. The task of deterministic entanglement
purification can be accomplished with two steps: one for correcting
the bit-flip errors and the other for the phase-flip errors. The two
parties in quantum communication, say Alice and Bob, can correct the
bit-flip errors completely in principle with the entanglement in the
spatial degree of freedom and they can correct the phase-flip errors
perfectly in principle with the entanglement in the frequency degree
of freedom. This two-step DEPP does not require that the photon
pairs have entanglement in the polarization degree of freedom.
Different from CEPPs, in which Alice and Bob can only improve the
fidelity of the remaining ensemble, they can obtain a maximally
entangled photon pair from each pair in hyperentanglement with this
two-step DEPP \cite{twostepepp}. However, the entanglement in two
degrees of freedom of photons increases the difficulty of its
implementation .

In fact, Simon and Pan \cite{Simon}   used  spatial entanglement to
purify the polarization entanglement in 2002. In the Simon-Pan
protocol \cite{Simon}, the two photons in each pair are entangled in
two degrees of freedom of photons; that is, they are hyperentangled
in the spatial and  polarization degrees of freedom. However, it is
only a CEPP. With a new decomposition method, the Simon-Pan protocol
can be improved to be a DEPP. In this Brief Report, we  present a
one-step DEPP with simple linear optics and a practical PDC source
by improving the Simon-Pan protocol. Alice and Bob can obtain the
maximally entangled pairs without  largely consuming the
less-entangled pairs but rather only the spatial entanglements by
postselection. Compared with the Simon-Pan protocol \cite{Simon},
this one-step DEPP has some advantages. First, it works in a
deterministic way, not a probabilistic one. That is, Alice and Bob
can obtain a maximally entangled state from each entangled photon
pair, which   reduces a great deal of entanglement resources
consumed in quantum communication. Second, it does not require
hyperentanglement in more than two degrees of freedom, nor  does it
require that the polarization state be entangled. Moreover, it is
feasible with current techniques as its implementation is not more
difficult than the experimental demonstration  shown in Ref.
\cite{Pan2}. All these advantages   make this one-step protocol more
convenient than other EPPs in the applications in quantum
communication.

Before we start the present DEPP, let us first explain the
generation of the spatial entanglement and polarization entanglement
with a PDC source. As shown in Fig.1, the pump pulse of  ultraviolet
light passes through a beta barium borate  crystal. Then a
correlated pair of photons will be produced with   probability $p$
in modes $a_{1}$ and $b_{1}$. The  pulse can also be reflected and
traverses the crystal a second time,   producing another correlated
pair into modes $a_{2}$ and $b_{2}$ with the same probability $p$.
One also  gets two pairs from one source or one pair from each
source with the same order of magnitude $p^{2}$. As $p\ll1$, one can
omit the probability $p^{2}$ because it is a second-order term. Then
we can describe approximately the PDC sources with the Hamiltonian
\cite{Simon,shengpra}
\begin{eqnarray}
H=\gamma[(a^{\dagger}_{1H}b^{\dagger}_{1H}+a^{\dagger}_{1V}b^{\dagger}_{1V})
+re^{i\theta}(a^{\dagger}_{2H}b^{\dagger}_{2H}+a^{\dagger}_{2V}b^{\dagger}_{2V})] \;\;\;\nonumber\\
+H.c,
\end{eqnarray}
where  subscripts $H$ and $V$   denote  horizontal and vertical
polarizations, respectively.  $r$ is the relative probability of
emission of photons into the lower modes ($a_{2}b_{2}$) compared to
the upper modes ($a_{1}b_{1}$), and $\theta$ is the phase between
these two possibilities. We can make $r=1$ and $\theta=0$ as a
simple case, the same as in the Simon-Pan protocol \cite{Simon}. One
can see that the item
$(a^{\dagger}_{1H}b^{\dagger}_{1H}+a^{\dagger}_{1V}b^{\dagger}_{1V}
+
a^{\dagger}_{2H}b^{\dagger}_{2H}+a^{\dagger}_{2V}b^{\dagger}_{2V})|0\rangle$
represents the entanglement in both the polarization and  the
spatial modes. So this single-pair state can also be written as
\begin{eqnarray}
|\Psi\rangle=\frac{1}{2}(|a_{1}\rangle|b_{1}\rangle+|a_{2}\rangle|b_{2}\rangle)(|H\rangle|H\rangle
+|V\rangle|V\rangle)_{ab}\label{state1}
\end{eqnarray}
in a different notation. The subscripts $a$ and $b$ represent the
two photons $a$ and $b$ sent to Alice and Bob, respectively. If we
denote $|\phi^{+}\rangle_{p}=\frac{1}{\sqrt{2}}(|H \rangle|H \rangle
+ |V \rangle|V\rangle)_{ab}$ and
$|\phi\rangle_{s}=\frac{1}{\sqrt{2}}(|a_{1}\rangle|b_{1}\rangle+|a_{2}\rangle|b_{2}\rangle)$,
Eq.(\ref{state1}) can also be rewritten as:
\begin{eqnarray}
\rho=\rho_{p}\otimes\rho_{s},\label{rho1}
\end{eqnarray}
where $\rho_{p}=|\phi^{+}\rangle_{p}\langle\phi^{+}|$ and
$\rho_{s}=|\phi\rangle_{s}\langle\phi|$.

Now we start our purification protocol. Its principle is shown in
Fig.1. If the two photons do not suffer from decoherence and remain
in the state with the form of Eq.(\ref{state1}), the whole state
evolves as:
\begin{eqnarray}
|\Psi\rangle & = &
\frac{1}{2}(|H_{a1}\rangle|H_{b1}\rangle+|V_{a1}\rangle|V_{b1}\rangle+|H_{a2}\rangle|H_{b2}\rangle
+|V_{a2}\rangle|V_{b2}\rangle)\nonumber\\
& \rightarrow &
\frac{1}{2}(|H_{c1}\rangle|H_{d1}\rangle+|V_{e1}\rangle|V_{f1}\rangle+|V_{c2}\rangle|V_{d2}\rangle
+|H_{e2}\rangle|H_{f2}\rangle).\nonumber\\
\end{eqnarray}
One can see that the items $|H_{c1}\rangle|H_{d1}\rangle$ and
$|V_{c2}\rangle|V_{d2}\rangle$ will be detected in  $D_{2}$ and
$D_{4}$, and  $|V_{e1}\rangle|V_{f1}\rangle$ and
$|H_{e2}\rangle|H_{f2}\rangle)$ will be detected in $D_{5}$ and
$D_{7}$. That is, if the coincidence detectors $D_{2}$ and $D_{4}$,
or $D_{5}$ and $D_{7}$ click (i.e., the two photons emit from the
outputs $D_2$ and $D_4$ or $D_5$ and $D_7$), Alice and Bob can get
the maximally entangled state $\vert
\phi^+\rangle_p=\frac{1}{\sqrt{2}}(|H\rangle|H\rangle +
|V\rangle|V\rangle)$ by postselection.

\begin{figure}[!h]
\includegraphics[width=8.4cm,angle=0]{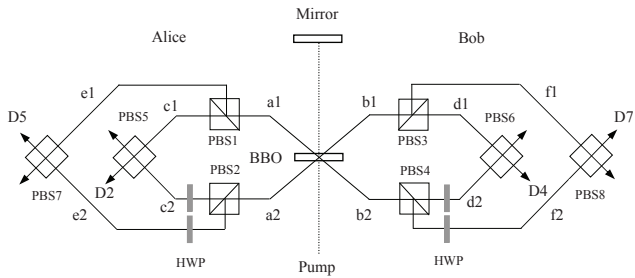}
\caption{Schematic illustration of the one-step deterministic
entanglement purification protocol. The two parametric
down-conversion sources are used to generate the photons being
entangled both in the polarization and  in the spatial-mode degrees
of freedom of photons. In the postselection after the polarizing
beam splitters (PBSs), the photon pairs have been purified in a
deterministic way.  The PBS is used to transmit the $|H\rangle$
polarization photons and reflect the $|V\rangle$ polarization
photons. The half-wave plates (HWPs) can convert the polarization
state  $|H\rangle$ into $|V\rangle$, or vice versa. Only four
detectors are used here. If the measurement results are $D_{2}$ and
$D_{4}$, or $D_{5}$ and $D_{7}$, the two photons are in the state
$|\phi^{+}\rangle$. If the measurement results are $D_{2}$ and
$D_{7}$ or $D_{4}$ and $D_{5}$, the two photons are in the state
$|\psi^{+}\rangle$. }
\end{figure}

So far we have been talking about the absence of noise in an ideal
case. In a practical transmission, the channel noise always exists
and   makes the pure state become a mixed one. The polarization part
of the entangled state may suffer from bit-flip and phase-flip
errors, and the phase between the modes $a_{1}b_{1}$ and
$a_{2}b_{2}$ is not exactly stable; that is, the spatial
entanglement will also be affected. Fortunately, by having them in
two different optical fibers, the bit-flip  errors of spatial
entanglement (i.e., the two spatial modes on each side) are
extremely small. Here we suppose that only the polarization
entanglement part suffers from noise, just as in the Simon-Pan
protocol \cite{Simon}. Through a noisy channel, Eq.(\ref{rho1})
becomes:
\begin{eqnarray}
\rho'=\rho_{p}'\otimes\rho_{s},\label{rhoallerror}
\end{eqnarray}
where
\begin{eqnarray}
\rho_{p}'&=&F|\phi^{+}\rangle_{p}\langle\phi^{+}|+F_{1}|\phi^{-}\rangle_{p}\langle\phi^{-}|\nonumber\\
&+& F_{2}|\psi^{+}\rangle_{p}\langle\psi^{+}|
+F_{3}|\psi^{-}\rangle_{p}\langle\psi^{-}|.\label{rho2}
\end{eqnarray}
Here $F+F_{1}+F_{2}+F_{3}=1$,
$|\phi^{-}\rangle_{p}=\frac{1}{\sqrt{2}}(|H \rangle|H \rangle-|V
\rangle|V\rangle)_{ab}$, and
$|\psi^{\pm}\rangle_{p}=\frac{1}{\sqrt{2}}(|H \rangle|V \rangle\pm|V
\rangle|H\rangle)_{ab}$. Based on the density matrix of
Eq.(\ref{rho2}), we denote that there are two kinds of errors in the
mixed state. One is the bit-flip error and the other is the
phase-flip error. For instance, if the state $|\phi^{+}\rangle_{p}$
becomes $|\psi^{+}\rangle_{p}$, we call it a bit-flip error. If
$|\phi^{+}\rangle_{p}$ becomes $|\phi^{-}\rangle_{p}$, a phase-flip
error occurs. $|\psi^{-}\rangle_{p}$ contains both a bit-flip error
and a phase-flip error. So the whole task of entanglement
purification is to correct these two kinds of errors.

If a bit-flip error takes place with a probability of $F_{2}$, after
the PBSs and half-wave plates (HWPs) shown in Fig.1, the state
becomes:
\begin{eqnarray}
&&\frac{1}{2}(|a_{1}\rangle|b_{1}\rangle +
|a_{2}\rangle|b_{2}\rangle)(|H \rangle|V \rangle
+|V \rangle|H\rangle)_{ab} \nonumber\\
&&=(|H_{a1}\rangle|V_{b1}\rangle+|H_{a2}\rangle|V_{b2}\rangle+|V_{a1}\rangle|H_{b1}\rangle
+|V_{a2}\rangle|H_{b2}\rangle) \nonumber\\
&&\rightarrow\frac{1}{2}(|H_{c1}\rangle|V_{f1}\rangle+|V_{c2}\rangle|H_{f2}\rangle
+|V_{e1}\rangle|H_{d1}\rangle+|H_{e2}\rangle|V_{d2}\rangle). \nonumber\\
\label{statebiterror}
\end{eqnarray}
We can also find that the items $|H_{c1}\rangle|V_{f1}\rangle$ and
$|V_{c2}\rangle|H_{f2}\rangle$ have the same outputs   $D_{2}$ and
$D_{7}$, and the items $|V_{e1}\rangle|H_{d1}\rangle$ and
$|H_{e2}\rangle|V_{d2}\rangle$ have the same outputs  $D_{4}$ and
$D_{5}$. That is, if the coincidence detectors $D_{2}$ and $D_{7}$
or $D_{4}$ and $D_{5}$ click, Alice and Bob can finally obtain the
maximally entangled state
$\frac{1}{\sqrt{2}}(|H\rangle|V\rangle+|V\rangle|H\rangle)$ by
postselection. Therefore, by performing a bit-flip operation
$\sigma_{x}=|H\rangle\langle V|+|V\rangle\langle H|$ on one of the
two photons, Alice and Bob can get rid of the bit-flip error and
then obtain the uncorrupted state $\vert
\phi^+\rangle_p=\frac{1}{\sqrt{2}}(|H\rangle|H\rangle+|V\rangle|V\rangle)$.

Following  the same principle discussed above, if the initial state
is $\frac{1}{2}(|H \rangle|H \rangle-|V
\rangle|V\rangle)_{ab}(|a_{1}\rangle|b_{1}\rangle+|a_{2}\rangle|b_{2}\rangle)$,
 in  other words, if a phase-flip occurs with a probability of $F_{1}$,
after the PBSs and the HWPs, the coincidence detectors will get the
results that $D_{2}$ and $D_{4}$, or $D_{5}$ and $D_{7}$ click. This
result is the same as the case with no phase-flip error [see
Eq.(\ref{rho1})]. If both a bit-flip error and a phase-flip error
occur, the initial state is $\frac{1}{2}(|H \rangle|V \rangle-|V
\rangle|H\rangle)_{ab}(|a_{1}\rangle|b_{1}\rangle+|a_{2}\rangle|b_{2}\rangle)$,
which has the same result as the case with only bit-flip errors [see
Eq.(\ref{statebiterror})]. Surprisingly, the phase-flip error has no
effect on the entanglement purification here, which is far different
from the CEPPs \cite{Bennett1,Deutsch,Pan1,Simon,shengpra}.

So far we have  described the principle of  our one-step DEPP. With
one-step DEPP, Alice and Bob can obtain the maximally entangled
state $\frac{1}{\sqrt{2}}(|H\rangle|H\rangle+|V\rangle|V\rangle)$ or
$\frac{1}{\sqrt{2}}(|H\rangle|V\rangle+|V\rangle|H\rangle)$,
corresponding to the coincidence detectors and two deterministic
output modes. That is, Alice and Bob can implement deterministic
entanglement purification by postselection. Compared with the
initial mixed state in Eq.(\ref{rhoallerror}), the spatial
entanglement has been consumed but polarization entanglement
remains.

The present one-step protocol for deterministic entanglement
purification is interesting because of its great simplicity and high
efficiency. In CEPPs \cite{Bennett1,Deutsch,Pan1}, Alice and Bob
should use two less-entangled copies to complete a round for
purification. After some local operations and classical
communication, one photon pair in a high-fidelity entangled state
will   remain  with some  probability. In this case, at last one of
the two less-entangled pairs is consumed. If we want to obtain
higher-quality  entangled pairs, more less-entangled pairs are
consumed. Surprisingly, in our one-step DEPP, the entanglement in
the polarization seems redundant and it does not require the
polarization part of the state be entangled; that is,  the
coefficients of $F$, $F_{1}$, $F_{2}$, and $F_{3}$ only need to
satisfy the condition $F+F_{1}+F_{2}+F_{3}=1$. But in the CEPPs
\cite{Bennett1,Deutsch,Pan1,Simon,shengpra}, the initial fidelity
$F$ should satisfy the condition $F>\frac{1}{2}$ to ensure the mixed
state is  entangled in the polarization degree of freedom. Because
local operations and classical communication can not increase
entanglement, entanglement purification is essentially the
transformation of the entanglement. In the conventional protocols
\cite{Bennett1,Pan1,Deutsch}, its transformation is completed
between the same kind of entanglement (i.e., polarization
entanglement) but with different particles. The Simon-Pan protocol
\cite{Simon} and the present one-step DEPP accomplish the
purification with the transformation between the  polarization
entanglement and the spatial entanglement. However, the spatial
entanglement in the Simon-Pan protocol \cite{Simon} is only used to
purify the bit-flip error. After consuming the entanglement in the
spatial degree of freedom, Alice and Bob have to adopt the CEPPs
\cite{Pan1,Simon} to improve the fidelity of the polarization
entanglement.  However,  in the QND protocol \cite{shengpra}, the
QND has the function of both a nondestructive single-photon detector
and a CNOT gate, and it also cannot reach the maximally entangled
pure state, and more less-entangled pairs will be consumed if  the
QND protocol is performed repeatedly.


To understand why this one-step DEPP has a success probability of
100\% for entanglement purification in principle, we employ another
example to explain this method. From the view of the outcome of the
measurement on the photon systems in the polarization degree of
freedom with a product basis, say $\sigma_z^A \otimes \sigma_z^B$,
the state of the polarization part after the transmission over noisy
channels can be described as:
\begin{eqnarray}
\rho_{p}'' &=& \alpha|HH\rangle\langle HH| + \beta|VV\rangle\langle
VV|\nonumber\\
&+& \gamma|HV\rangle\langle HV| + \delta|VH\rangle\langle VH|,
\end{eqnarray}
where $\alpha$, $\beta$, $\gamma$, and $\delta$ represent the
probabilities that Alice and Bob obtain the product states
$|HH\rangle$, $|HV\rangle$, $|VH\rangle$, and $|VV\rangle$,
respectively, when they measure their photon pair with the basis
$\sigma_z^A \otimes \sigma_z^B$. $\alpha^2+ \beta^2 + \gamma^2 +
\delta^2=1$. That is,  $\rho'$ can be rewritten as:
\begin{eqnarray}
\rho'=\rho_{p}''\otimes\rho_{s}.\label{rhoproduct}
\end{eqnarray}
So $\rho'$ can be viewed as a probabilistic mixture of four pure
states; this pair is in the state
$|HH\rangle\otimes(|a_{1}\rangle|b_{1}\rangle+|a_{2}\rangle|b_{2}\rangle)$,
$|VV\rangle\otimes(|a_{1}\rangle|b_{1}\rangle+|a_{2}\rangle|b_{2}\rangle)$,
$|HV\rangle\otimes(|a_{1}\rangle|b_{1}\rangle+|a_{2}\rangle|b_{2}\rangle)$,
or
$|VH\rangle\otimes(|a_{1}\rangle|b_{1}\rangle+|a_{2}\rangle|b_{2}\rangle)$
with the probability of $\alpha$,  $\beta$,  $\gamma$, or $\delta$,
respectively.


Following the same principle above,
$|HH\rangle\otimes(|a_{1}\rangle|b_{1}\rangle+|a_{2}\rangle|b_{2}\rangle)$
  leads to the two-mode case in the output $D_{2}$ and $D_{4}$,
and becomes the maximally entangled state
$\frac{1}{\sqrt{2}}(|H\rangle|H\rangle+|V\rangle|V\rangle)$.
$|VV\rangle\otimes(|a_{1}\rangle|b_{1}\rangle+|a_{2}\rangle|b_{2}\rangle)$
 has the same result above but leads to the output  $D_{5}$ and $D_{7}$.
$|HV\rangle\otimes(|a_{1}\rangle|b_{1}\rangle+|a_{2}\rangle|b_{2}\rangle)$
is in the output $D_{2}$ and $D_{7}$ with the remaining  state
$\frac{1}{\sqrt{2}}(|H\rangle|V\rangle+|V\rangle|H\rangle)$, and
$|VH\rangle\otimes(|a_{1}\rangle|b_{1}\rangle+|a_{2}\rangle|b_{2}\rangle)$
will have the same result but leads to the output  $D_{4}$ and
$D_{5}$. We know that in each item, for instance, the state
$|HH\rangle\otimes(|a_{1}\rangle|b_{1}\rangle+|a_{2}\rangle|b_{2}\rangle)$,
there is no entanglement in polarization, but the spatial part is
entangled. After purification, the spatial entanglement has been
transformed into the polarization degree of freedom successfully.

It is interesting to compare this protocol with the Simon-Pan
protocol \cite{Simon}. In their protocol, the spatial entanglement
is used to purify polarization entanglement, by selecting  both
upper and lower modes, and then Alice and Bob can correct the
bit-flip error successfully. However, the phase-flip error can not
be purified directly. They have to use the CEPPs
\cite{Bennett1,Deutsch} to get rid of it. It is well known that a
mixed state can be decomposed in any orthogonal basis. For example,
like Eq.(\ref{rhoallerror}), it is decomposed in a Bell-state basis,
but in Eq.(\ref{rhoproduct}) it is in a product-state basis. In
Eq.(\ref{rhoallerror}), there are two kinds of errors, bit-flip
error and phase-flip error, but no phase-flip error exists in
Eq.(\ref{rhoproduct}). The state may be one bit-flip error, like
$|HV\rangle$ or $|VH\rangle$, or two bit-flip errors, like
$|VV\rangle$. In this one-step DEPP, the initial mixed state has
been first divided into different bit-flip errors ($|HH\rangle$,
$|VV\rangle$, $|HV\rangle$, or $|VH\rangle$) and guided into
different spatial modes with $PBS_{1}$, $PBS_{2}$, $PBS_{3}$, and
$PBS_{4}$. Then another four PBSs are used to convert the spatial
entanglement into the polarization entanglement perfectly by
postselection. This is the reason that
$|\phi^{+}\rangle_{p}\otimes|\phi\rangle_{s}$ and
$|\phi^{-}\rangle_{p}\otimes|\phi\rangle_{s}$ have the same outputs
and the phase-flip error has been eliminated automatically. However,
in the Simon-Pan protocol \cite{Simon}, the two PBSs can only be
used to discriminate the bit-flip error and the phase-flip error
remains. So the transformation efficiency is only 50\%. Moreover,
after the single-photon detector, the fidelity can not be increased
any more because the spatial entanglement is consumed completely.

In summary, we have presented  a one-step deterministic entanglement
purification protocol with simple linear optical elements. Compared
with other protocols, this protocol has several advantages. First,
it can obtain a maximally entangled pair with only one step, instead
of improving the fidelity of less-entangled pairs by performing the
purification protocol repeatedly. Second, Alice and Bob do not need
 many less-entangled pairs because this one-step protocol works in
a deterministic way, not a probabilistic one. This one-step protocol
can greatly reduce  the number of entanglement resources needed.
Third, it does not require that the polarization state be entangled;
only spatial entanglement is needed. The most important advantage of
this one-step DEPP may be its realization with current technology.
In a previous experiment \cite{Pan2}, the main experimental
requirements of good mode overlap on the PBSs and phase stability in
the spatial mode have both been achieved. All these advantages  make
this one-step protocol more convenient than others
\cite{Bennett1,Deutsch,Pan1,Simon,shengpra,twostepepp} in current
quantum communication.

Although we only discuss the present one-step DEPP with spatial
entanglement, it works with frequency entanglement if the latter
suffers less from the channel noise. That is, the entanglement
transformation can also be accomplished between the frequency degree
of freedom and the polarization degree of freedom, and the method
for the  decomposition of quantum states in this one-step DEPP is
general for other degrees of freedom of photons.


This work is supported by the National Natural Science Foundation of
China under Grant No. 10974020, a Foundation for the Author of
National Excellent Doctoral Dissertation of P. R. China under Grant
No. 200723, and the Beijing Natural Science Foundation under Grant
No. 1082008.

\end{document}